# A Proposed NFC Payment Application

Pardis Pourghomi
School of Information System,
Computing and Mathematics
Brunel University
London, UK

Muhammad Qasim Saeed
Information Security Group (ISG)
Royal Holloway University of
London Egham, UK

Gheorghita Ghinea
School of Information System,
Computing and Mathematics
Brunel University
London, UK

*Abstract*—Near Field Communication (NFC) technology is based on a short range radio communication channel which enables users to exchange data between devices. With NFC technology, mobile services establish a contactless transaction system to make the payment methods easier for people. Although NFC mobile services have great potential for growth, they have raised several issues which have concerned the researches and prevented the adoption of this technology within societies. Reorganizing and describing what is required for the success of this technology have motivated us to extend the current NFC ecosystem models to accelerate the development of this business area. In this paper, we introduce a new NFC payment application, which is based on our previous "NFC Cloud Wallet" model [1] to demonstrate a reliable structure of NFC ecosystem. We also describe the step by step execution of the proposed protocol in order to carefully analyse the payment application and our main focus will be on the Mobile Network Operator (MNO) as the main player within the ecosystem.

## I. INTRODUCTION

During the past decade, the concept of contactless card technology was introduced to be used in transport, ticketing and in retail. The technology helps people save time by just holding their contactless cards against a reader in a close proximity instead of having to insert the paper cards in and taking it out of that entrance gates for example. With NFC technology, mobile phones can have additional functionality to act as a contactless card to be used as an easy method of payment. Successful development of NFC technology has recently started in some countries where companies offer several services based on the contactless card technology and mobile phones. Although this technology is increasingly becoming mainstream, it still has issues that need to be addressed [2]. These issues are mainly security concerns with Secure Element (SE) personalization, management, ownership and architecture that can be exploitable by attackers to delay the adaption of NFC within societies. The purpose of this paper is to extend Pourghomi's[1] - this model will be referred as "NFC Cloud Wallet" in this and our future papers - by proposing a complete transaction mechanism based on NFC and GSM networks.

This model is based on cloud computing for the management of payment applications in secure element within the NFC ecosystem. The details of this model are described in Section IV of this paper. As the authentication mechanism of our extended model is based on GSM, we will discuss the GSM authentication later in this paper. We also aim to accelerate the development of NFC mobile payment services by describing the NFC ecosystem in order to raise the attention of business players in terms of the new potential models that can be implemented in order to achieve a cost beneficial and less complex ecosystem framework.

Our contribution in this paper is to extend the NFC Cloud Wallet model and to provide a complete transaction solution based on this model. We propose a model based on the assumption that the cloud is being managed by the MNO.We used the existing security features of GSM network to achieve authentication, data integrity and data confidentiality.

In our proposed model, the SIM is the secure element which is being managed by the MNO. By using our model, a customer with a NFC enabled mobile phone can pay through his cell phone in a secure way.

This paper is organized as follows. Section II consists of a brief introduction to NFC ecosystem with its main elements and functionalities. Section III describes the roles of Secure Element (SE) and the Universal Integrated Circuit Card (UICC) within the NFC ecosystem. Section IV evaluates the previously proposed NFC Cloud Wallet model. Section V discusses GSM authentication that is used in our extended model. Section VI introduces the proposed transaction model as well as the proposed transaction protocoland describes its execution process in details. Section VII is the analysis of our proposed model from multiple security aspects. This analysis encompasses the authentication and security of the messages among customer, shop POS terminal and the MNO. Finally, Section VIII presents our conclusion.

## II. NFC & NFC ECOSYSTEM

This section describes the functions of adding the contactless card to mobile phones which produce an intelligent device that enables us to make payments with. This intelligent device is called a "NFC Mobile Phone". When different functions of a mobile phone combine with the functions of contact-less cards, the results of this combination will have a greater significance than just the importance of adding two devices together. This significance defines the NFC-enabled mobile phone which can connect with another NFC-enabled device (i.e. PDA, tablets, etc.) in a short range communication channel. NFC technology enables users to benefit from new and countless services on a daily basis where they can pay for their food; buy a cinema ticket by scanning their phone against a movie poster and much more. This newly developed intelligent device is proposed as an all-in-one personal device that can be personalized and used in a highly interactive environment [3]. Fig. 1 demonstrates the concept of the NFC mobile phone [4].

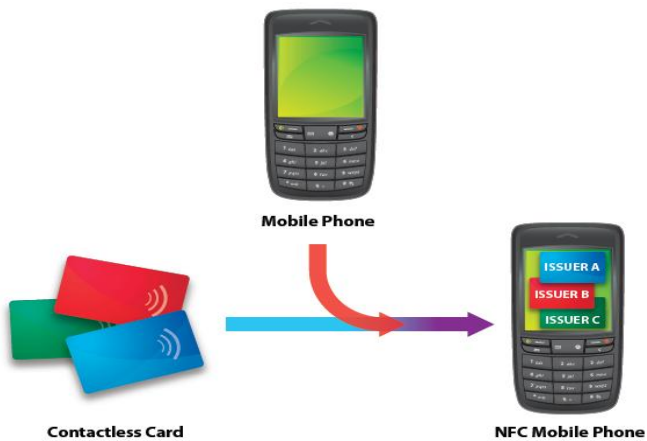

**Mobile Phone**

**Contactless Card**

**NFC Mobile Phone**

ISSUER A
ISSUER B
ISSUER C

Fig. 1. The concept of NFC mobile phone

The success of the NFC mobile ecosystem is based on the relationships between the involved parties where those relationships have to be clearly defined. The present contactless ecosystem models functionalities can be extended by a well-defined NFC ecosystem which improves the number of functionalities that an involved party can provide. Table I describes the key functionalities of NFC ecosystem [4].

TABLE I. KEY FUNCTIONALITIES OF NFC ECOSYSTEM

| Key functionalities | Description |
|---|---|
| Service provisioning | It provides authentication and remote user management due to network availability. Also users can subscribe and personalize their contactless cards. |
| Mobile network provisioning | It offers user authentication and user care for data connectivity as well as ensuring that the network infrastructure is maintained to enable users to receive data connectivity service. |
| Trusted Service Manager (TSM) | Delivers a communication platform between Service Providers (SPs) and NFC mobile phones where SPs provide multi-application management functionalities to NFC enabled mobile phone through this platform. |

### III. SECURE ELEMENT (SE)

The security of NFC is supposed to be provided by a component called security controller that is in the form of a SE. The SE is an attack resistant microcontroller more or less like a chip that can be found in a smart card [3].

SE provides storage within the mobile phone and it contains hardware, software, protocols and interfaces. SE provides a secure area for the protection of the payment assets (e.g. keys, payment application code, and payment data) and the execution of other applications. In addition, SE can be used to store other applications which require security mechanisms and it can also be involved in authentication processes. To be able to handle all these, the installed operating system has to have the capability of personalizing and managing multiple applications that are provided by multiple SPs preferably Over-The-Air (OTA). Still the ownership and control of SE within NFC ecosystem may result in a commercial and strategic advantage but some solutions are already in place and researchers are developing new models to overcome this problem. Universal Integrated Circuit Card is (UICC) is one of the most reliable components to act as a SE in NFC architecture [5]. It is removable, provides the same security as a smartcard, can run multiple applications issued by multiple providers, it is compliant with all smart card standards and it supports GSM and UMTS network. According to GSMA guidelines, UICC is the most appropriate NFC Secure Element in mobile phones [3].

#### A. SE Lifecycle

The **Initialization** of an SE can be completed by different SE issuers such as credit card companies, Mobile Network Operator (MNO), financial institution or retailers. The SEI can also act as a platform provider. If the SE does not contain any applications when issued that means there is no platform manager assigned to that SE. A platform manager cannot deal with SE applications without having different certifications (i.e. Visa PayWave certification).

The *Activation* process takes place when the SE is inserted into the phone. The SE then sings in to the NFC controller and NFC controller sends a confirmation message to the platform manager in order to inform the platform manager of the successful insertion of the SE in the phone. The platform manager then sends a confirmation message to the mobile phone in order to activate the SE. The platform manager is the only party that has the authority to hold the SE keys for data configuration purposes. NFC controller's identifier is also stored in the SE to inform SE in case if it was inserted into another phone.

During phase 1 of the *Applications Upload* process, the SP (in this case also the Application Issuer (AI)) contacts the MNO that is the only party who is in charge of the Mobile Station International ISDN Number (MSISDN). The only way to classify the external party for an Over-The-Air (OTA) transaction with the NFC phone is the MSISDN.

In phase 2, MNO forwards the SP request to the platform manager (s) that is in charge of the SE. If the there is no SE in the phone, the MNO will inform the SP regarding this issue.

In this case application upload process terminates. But if the platform manager is positive with the request, it will send an offer directly to the SP to upload its application.

In the next phase, SP selects one platform manager amongst others (if more than one platform manager exists) to load its data to the security domain area which is under the control of the same platform manager.

The *Deactivation* procedures are also managed by the platform manager where it can deactivate the SE, OTA in the case of theft or loss. If SE is installed in a new device, then the activation process should be renewed and the platform manager is the only party that should confirm the activation process to enable the SE to be used for contactless transactions [5].

## IV. NFC Cloud Wallet Model

This model brought the idea of using cloud computing in order to manage the NFC payment applications which resulted in flexible and secure management, personalization and ownership of the applications [1].This architecture provides easy management of multiple users and delivers personalized contents to each user. It supports intelligent profiling functions by managing customized information relevant to each user in certain environments which updates the service offers and user profiles dynamically. Depending on the MNO network's reception, deployment of this service takes around one minute and deployments can be scaled to any number of users.

The idea of this approach is that every time the customer makes a purchase the payment application which contains the customer's credentials is downloaded into the mobile device (SE) from the cloud and, after the transaction, it is deleted from the device and the cloud will update itself to keep a correct record of customer's account balance. Fig. 2 illustrates the steps that should be undertaken to complete the transaction process [1].

The execution of the model is described in what follows:

*1) Customer waves the NFC enabled phone on the POS terminal to make the payment*

*2) The payment application is downloaded into customer's mobile phone SE.*

*3) The reader communicates with the cloud provider to check whether the customer has enough credit or not.*

*4) Cloud provider transfers the required information to the reader.*

*5) Based on the information which was transferred to the reader, the reader either authorizes the transaction or rejects customer's request.*

*6) Reader communicates with the cloud to update customer's balance - if customer's request was authorised, the amount of purchase will be withdrawn from his account otherwise customer's account will remain with the same balance.*

As an addition to this model, we suggest that when the NFC enabled phone sends a request to its cloud provider to get permission to make a payment (step 1), the cloud provider sends a SMS requesting a PIN number to identify the user of the phone - this is how cloud provider ensures the legitimacy of the phone user. For verification purposes, the customer sends the PIN back to the cloud provider as an SMS.

In order to extend this model, there are two possible approaches to follow. Firstly, the financial institution can be the cloud owner from which the payment application can be downloaded from/into the customer's mobile device; MNO can be linked to the financial institution (that is the cloud owner in this case) or it can stand as a separate party. Secondly, the financial institution could have a contract with a third party company such as PayPal that has its own cloud infrastructure (MNO can be linked with them, it also can stand as a separate party) or the financial institution uses other company's cloud service such as IBM, Microsoft, etc (MNO can be linked with either financial institution, cloud provider or it can stand as a separate party).

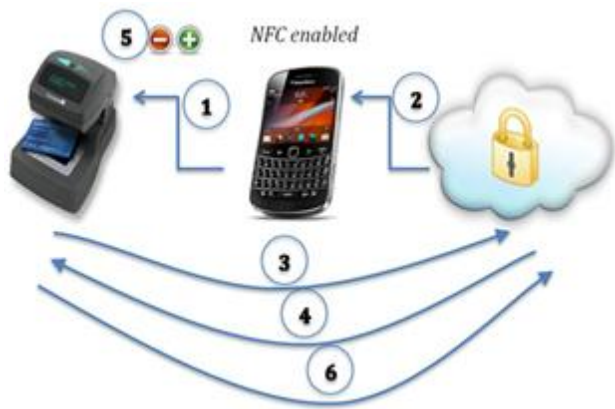

Fig. 2. NFC cloud wallet

## V. GSM Authentication

When a mobile device signs into a network, the Mobile Network Operator (MNO) first authenticates the device (specifically the SIM). The authentication stage verifies the identity and validity of the SIM and ensures that the subscriber has authorized access to the network. The Authentication Centre (AuC) of the MNO is responsible for authenticating each SIM that attempts to connect to the GSM core network through Mobile Switching Centre (MSC). The AuC stores two encryption algorithms A3 and A8, as well as a list of all subscribers identity along with corresponding secret key $K_i$.

This key is also stored in the SIM. The AuC first generates a random number known as $R$. This R is used to generate two responses, signed response $S$ and key $K_c$ as shown in Fig. 3, where $S = E_{Ki}(R)$ using A3 algorithm and $K_c = E_{Ki}(R)$ using A8 algorithm [6][7][8][9].

The triplet $(R, S, K_c)$ is known as Authentication triplet generated by AuC. AuC sends this triplet to MSC. On receiving a triplet from AuC, MSC sends $R$ (first part of the triplet) to the mobile device. SIM of the mobile device computes the response $S$ from $R$, as $K_i$ is already stored in the SIM. Mobile device transmits S to MSC. If this $S$ matches the $S$ in the triplet (which it should in case of a valid SIM), then the mobile is authenticated.$K_c$ is used for communication encryption between the mobile station and the MNO. Table IIdescribes the abbreviations used in the proposed protocol.

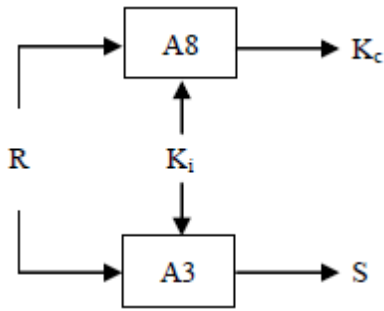

Fig. 3.   Generation of $Kc$ and $S$ from $R$

TABLE II.        ABBREVIATIONS

| | |
|---|---|
| $AuC$ | Authentication Centre (subsystem of MNO) |
| $IMSI$ | Internet Mobile Subscriber Identity |
| $K_i$ | SIM specific key. Stored at a secure location in SIM and at AuC |
| $K_c$ | $E_{ki}(R)$ using A8 algorithm |
| $K_{c1}$ | $H(K_c)$. Used for MAC calculation |
| $K_{c2}$ | $H(K_c)$. Encryption key |
| $K_p$ | Shared key between PG and shop POS terminal |
| $LAI$ | Local Area Identifier |
| $MNO$ | Mobile Network Operator |
| $NFC$ | Near Field Communication |
| $PI$ | Payment Information |
| $POS$ | Point Of Sale. Part of shop |
| $R$ | $RAND$. Random Number (128 bits) |
| $R_s$ | Random number generated by SIM (128 bits) |
| $TC$ | Transaction Counter |
| $TRM$ | Transaction Request Message |
| $TI$ | Transaction Information |
| $TMSI$ | Temporary Mobile Subscriber Identity |
| $TP$ | Total Price |
| $TS_U$ | User's Time Stamp |
| $TS_{Tr}$ | Transaction Time Stamp |
| $SD$ | Shopping Details |

## VI.   PROPOSED MODEL

We propose an extension to previously proposed NFC

Cloud Wallet model. Since there are multiple options applicable to this model, we designed our model based on the following assumptions:

- SE is part of SIM
- Cloud is part of MNO
- MNO is managing SE/SIM
- Banks, etc. are linked with MNO

These assumptions are appropriate regarding the NFC execution process and its ecosystem. As mentioned in Section IIIpreviously, SE is in the format of UICC therefore SE is part of the SIM. MNO manages the cloud infrastructure and it is the only party that has full access and permission to manage confidential data which are stored in the cloud. As MNO is the owner of the cloud, it fully manages the SIM in terms of monitoring the GSM network and controlling cloud's data. From the financial institution's point of view, they only deal with MNO as MNO is the single party that has full control over the SIM as well as the cloud.

### A.  The Proposed Protocol

Our proposal is based on cloud architecture where the cloud is being managed by the MNO. The cloud and the banking sector are the subsystems of MNO in our proposal, in addition to the existing subsystems of an MNO. We assume that the communication is secure between various subsystems of the MNO. The shop POS terminal, registered with one or more MNO, shares an MNO specific secret key $K_p$ with the corresponding MNO. This key is issued once a shop is registered with the MNO. The bank detail of the shopkeeper is also registered with the MNO for monetary transactions. The communication between the shop POS terminal and the mobile device is wireless using NFC technology. The mobile device has a valid SIM. We used the existing feature of GSM network for mutual authentication. A recent study by reference [10] proposed a mechanism for GSM authentication in NFC environment. We tailored their model according to our requirement in our proposed architecture. The detailed execution of our protocol is described in Fig. 4.

The proposed protocol executes in three different phases: Authentication, Keys generation and Transaction. The protocol initiates when the customer places his cell phone for the payment after agreeing to the total price displayed on the shop POS terminal. The details of these phases are described in what follows:

### B.  Phase 1. Authentication

**Step 1:** As soon as the user places his mobile device, NFC link between the mobile device and the shop POS terminal is established. The shop POS terminal sends an *ID* Request message to the mobile device.

**Step 2-3:** The mobile device sends *TMSI*, *LA*I as its*ID*. On receipt of the information from the mobile device, the shop POS terminal determines the user's mobile network. The network code is available in *LAI* in the form of Mobile Country Code (*MCC*) and Mobile Network Code (*MNC*). An *MNC* is used in combination with *MCC* (also known as a *'MCC/MNC tuple'*) to uniquely identify a mobile phone operator/carrier [11].

**Step 4-5:** The shop POS terminal sends *TMSI*, *LAI*, and Shop *ID* to respective MNO for customer authentication and shop identification.

**Step 5.1:** In case of incorrect *TMSI*, a declined message is sent.

**Step 6:** In case of correct identification, the MNO generates one set of authentication triplet ($R$, $S$, $K_c$) and sends $R$ to mobile device through shop POS terminal.

**Step 7-8:** SIM computes $K_c$ from $R$ as explained in Section V. SIM generates a random number $R_s$ and concatenates with $R$, encrypts with key $K_c$ and sends it to the MNO through shop POS terminal.

**Step 9-10:** The MNO checks the validity of the SIM (or mobile device). It receives $E_{Kc}(R\|R_s)$ from the mobile device and decrypts the message by $K_c$, the key it already has in authentication triplet. The MNO compares $R$ in the authentication triplet with the $R$ in the response. In case they do not match, a 'Stop' message is sent to the mobile device and the protocol execution is stopped. If both $R$ are same, then the mobile is authenticated for a valid SIM. In this case, the MNO swaps $R$ and $R_s$, encrypts with $K_c$ and sends it to mobile device.

**Step 11-12:** This step authenticates the MNO to the mobile device. The mobile device receives the response $E_{Kc}$ ($R_s\|R$) and decrypts it with the key $K_c$ already computed in Step 7. The mobile device compares both $R$ and $R_s$. If both are same, then the MNO is authenticated and a *'successful authentication'* message is sent to the MNO.

### C. Phase 2. Key Generation and PIN Verification

**Step 13-14:** $K_p$ is a shared secret between the MNO and the shop POS terminal. $K_c$ is the shared secret between the MNO and the customer's mobile device (computed in step 7). There is no shared secret between the POS terminal and the mobile device till this stage. MNO and mobile device compute one-way hash function of $K_c$ to generate $K_{c1}$, the key that will be used for MAC calculation. The MNO computes $K_{c2}$ from $K_{c1}$ using one-way hash function and sends it to shop POS terminal by encrypting it with $K_p$. Mobile device also computes $K_{c2}$ as it already has $K_{c1}$. $K_{c2}$ is the encryption key between MNO, shop POS terminal and the customer's mobile device.

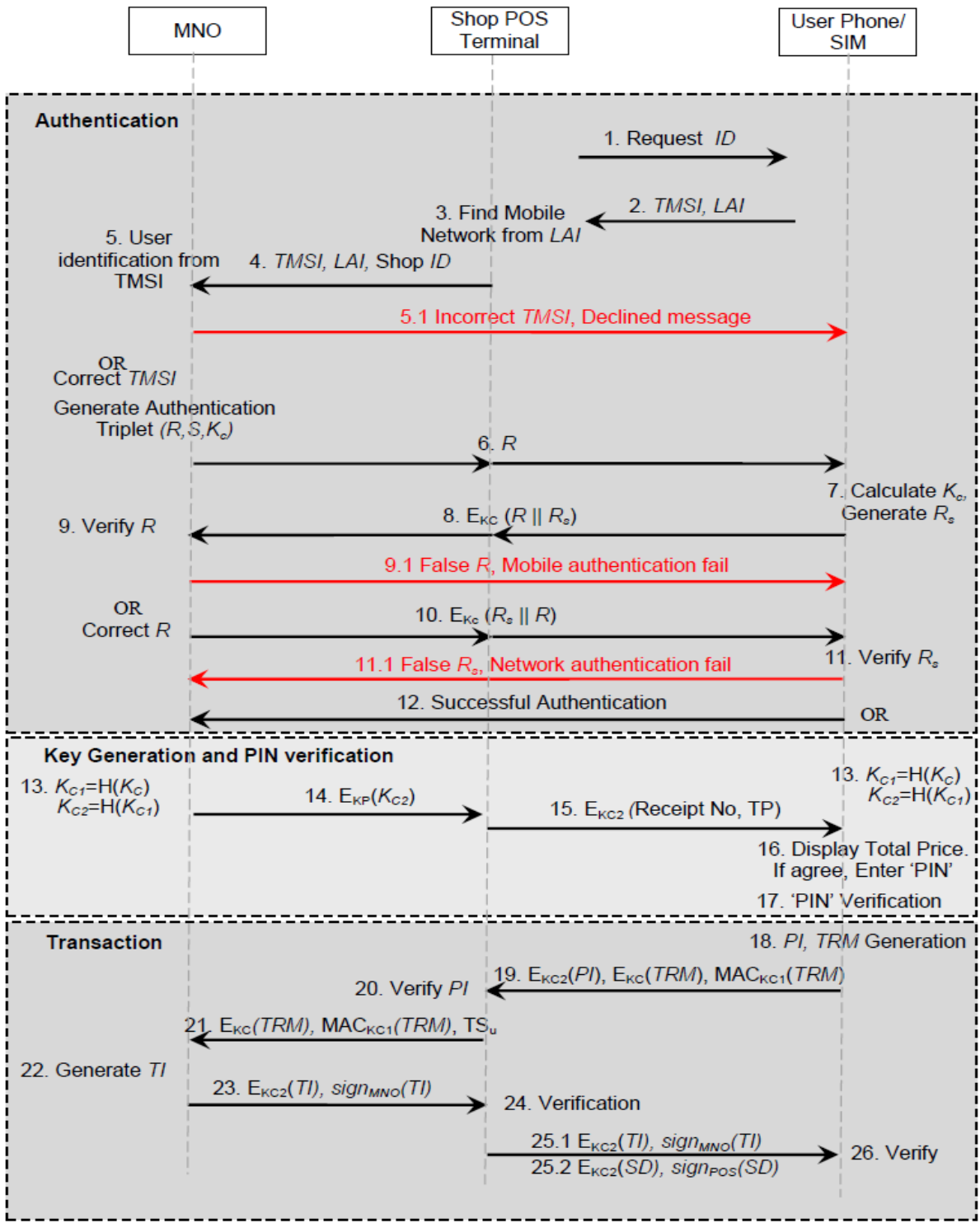

Fig. 4.   The proposed protocol

**Step 15-17:** The shop POS terminal sends the Total Price (*TP*) and the Receipt Number encrypted with $K_{c2}$. The user's mobile device decrypts the information and displays to the user. If he agrees, he enters the PIN. The PIN is an additional layer of security and adds trust between the user and the shopkeeper. A PIN binds a user with his mobile device, so the shopkeeper is to believe that the user is the legitimate owner of the mobile device. Moreover, the user feels more secure as no one else can use his mobile device for transaction without his consent. PIN is stored in a secure location in the SIM. The SIM compares both PINs and if both are same, the user is authenticated as the legitimate user of the mobile device. Otherwise, the protocol is stopped.

*D. Phase 3. Transaction*

**Step 18:** The customer's cell phone generates two messages, *PI* and *TRM*, such that;

$PI$= Receipt No, Total Price, Time Stamp ($TS_U$)
$TRM$=PI, $R_s$, Transaction Counter

**Step 19:** $TS_U$ represents the exact time and date the transaction has been committed by the user. *TC* is a counter that is incremented after each transaction and is used to prevent replay attack. *PI* is encrypted with $K_{c2}$ so that it can be verified by the shop POS terminal. The user encrypts the *TRM* with $K_c$ so that it cannot be modified by the shop terminal. The user computes MAC with $K_{cI}$ over the *TRM* using Encrypt-then-MAC approach for integrity protection.

**Step 20-21:** The POS terminal can decrypt only the *PI* encrypted with by $K_{c2}$ to check its correctness. The POS terminal does not need to verify the MAC (and it cannot do so), as it already knows the main contents of *PI*. The Shop POS terminal also verifies the $TS_U$ to be in a defined time window. If *PI* is correct, the POS terminal relays the encrypted *TRM* with corresponding MAC along with the $TS_U$ to the MNO.

**Step 22:** On receipt of the message, the MNO checks the integrity of the message by verifying the MAC with $K_{cI}$. If the MAC is invalid, the transaction execution is stopped. In case of a valid MAC, the MNO decrypts the message. The MNO compares the $R_s$ in the *TRM* with the $Rs$ received earlier in the authentication phase. A correct match confirms that the user is the same who was earlier authenticated. It also verifies the *TC* and $TS_U$. In case of successful verification, the MNO communicates with the concerned subsections for monetary transaction. The concerned subsections of the MNO checks the credit limitations of the user, and if satisfied, executes the transaction. Once the transaction is executed, the MNO generates *Transaction Information (TI)* message as:

*TI = Transaction Serial Number, Amount, $TS_{Tr}$*

**Step 23-25:** The MNO encrypts *TI* with $K_{c2}$, digitally signs the message and sends it to the shop POS terminal. The POS terminal verifies the signature. A valid signature indicates correct *TI*. The POS also verifies the *TI* for the amount mentioned in the *TI*. In case of successful verification, the POS terminal appends the message it received from the MNO with the *Shopping Details (SD)* and corresponding digital signature.

**Step 26:** The user verifies both signatures. It verifies the contents of *TI* and *SD*.

In this section, we analyse our proposed model from multiple security aspects. This analysis encompasses the authentication and security of the messages among customer, shop POS terminal and the MNO. The analysis also includes multiple attack scenarios, such as a customer is dishonest and has intentions to pay less, or the shopkeeper is dishonest and has plans to receive more money.

*A. Mutual Authentication*

A mutual authentication between a customer and MNO occurs whenever the customer agrees to pay some amount. Since this authentication is performed through shop POS terminal, we analysed our protocol from an angle that if the POS terminal has some malicious intentions. In this case, there can be following two scenarios:

*1) POS Terminal Impersonation as a Customer*
We assume that the shop POS terminal is dishonest and keeps a record of all messages against a legitimate customer (we call it as 'target scenario'). The aim of the shopkeeper is to transfer money from the target customer without his consent. The shop POS terminal impersonates as target customer to the MNO by replaying message 4. In case the *TMSI* and *LAI* are valid at that time (the chances are higher if the message is replayed just after the legitimate transaction of the target customer), the MNO will send a random number *R* to the terminal. *R* is 128 bit random number generated by the MNO so the chances for its repetition are almost negligible. The shopkeeper cannot compute a valid response in step 8 for a different *R*, as the shop lacks $K_i$ to compute $K_c$. Therefore, a shop cannot successfully impersonate as a customer by replaying old messages.

*2) POS Terminal Impersonation as MNO*
In this scenario, we assume that the shop is dishonest and communicates with a target customer without establishing a communication link with the MNO. Again, we assume that the shop keeps a record of legitimate messages of the target customer. The shop sends message 1 (*Request ID*) to the target customer and gets its response in message 2. Since shop does not communicate with MNO in this scenario, it does not send message 4 to MNO. However, the shop replays the recorded *R* in message 6 to the target customer. The target customer believes that he has been correctly identified by the MNO and the *R* is actually generated the MNO. So the user computes a response and sends it in message 8 to the shop. Message 8 contains $R_s$ encrypted with the $K_c$. The $R_s$ is a random number generated by the SIM and is different in each transaction. So, message 8 will be different than the one already recorded with the shop. Since message 8 is different, the shop can neither replay message 10, as it will be different for this transaction, nor it can compute a valid message 10. This scenario is, again, not successful.

### B. Encryption and MAC Keys

Separate keys are used for encryption and MAC calculation making the protocol more secure. *Encrypt-then-MAC* is an approach where the ciphertext is generated by encrypting the plaintext and then appending a MAC of the encrypted plaintext. This approach is cryptographically more secure than other approaches [12]. Apart from cryptographic advantage, the MAC can be verified without performing decryption. So, if the MAC is invalid for a message, the message is discarded without decryption. This results in computational efficiency.

### C. User Interaction

The user interaction with the system is reduced to single interaction making it a user-friendly protocol. The user feels more secure as the transaction is protected by PIN verification. There are chances that a user withdraws his mobile device from NFC terminal as a psychological move to enter PIN. This will break NFC link, but as the PIN is stored in the SIM, it does not require NFC link for verification. Once the user PIN has been verified by the SIM, the user places his mobile device back on the NFC terminal and the protocol resume from the same point. There are chances that a dishonest user withdraws his mobile device in order to enter the PIN, and then places back another mobile device for transaction. To counter this threat, $R_s$ is transmitted by the mobile device in Transaction Request Message (message 19). $R_s$ is generated by the SIM and is encrypted with $K_c$(message 7, 8), so it cannot be eavesdropped in the authentication phase. This ensures that the mobile device does not change.

### D. Disclosure of relevant Information

The protocol is designed considering disclosure of information on a need to know basis. For example, *TC* is a counter that increments after each successful transaction. The record of the *TC* is kept by both, the user and the MNO. Shop POS terminal does not need to know the *TC*. In our proposed protocol, the *TC* is not exposed to POS terminal as it is a part of *TRM*. Similarly, the MNO does not need to know the shopping details of the customer. Therefore, only the total amount is transmitted to the MNO for transaction.

### E. Transaction security

The transaction phase of the protocol requires maximum security. The *TRM* message is initiated by the customer rather than the shop terminal in order to satisfy the customer. The integrity of the *TRM* message is protected by the MAC so any alteration in this message is not possible. The message 19 is designed in such a way that the first half of the message containing encrypted *PI* is for shop POS terminal. POS terminal can decrypt and check the authenticity of the payment information. The remaining half of the message, containing encrypted *TRM* and corresponding MAC can neither be decrypted nor altered by the shop POS terminal. The POS terminal relays the remaining half to the MNO along with the Time Stamp. Hence, the transaction information generated by the customer is relayed to the MNO without any alteration.

In this phase, there can be a scenario where a dishonest customer has an intention of paying less than the actual amount. The customer designs a malicious *TRM* message (*TRM´*) consisting of *PI´* (an illegitimate payment information, *PI´<PI*). The dishonest customer then forms message 19 as:

$$PI = \text{Receipt Number, Total Price, Time Stamp } TS_U$$
$$TRM´ = PI´, R_s, \text{Transaction Counter (TC)}$$

It may be noted that the *PI* is legitimate whereas, the *TRM´* consists of amended *PI* (*PI´*). The dishonest customer forms message 19 as:

$$Message\ 19 = E_{kc2}(PI),\ E_{Kc}\ (TRM´),\ MAC_{Kc1}\ (TRM´)$$

The first half of the message, consisting of encrypted *PI*, is legitimate and the shop can verify it. However, the malicious part cannot be decrypted by the shop, so the shop cannot determine that the remaining part contains amended price information. The shop forms message 21 as *PI* is verified. The MNO executes the transaction with amended price and forms message 23 and digitally signs it. Message 23 contains the information about the amount deducted from the customer. Once this message is received by the shop terminal, the shop detects that the deducted amount is not the same as required. Hence, a dishonest customer with the intention to pay a lesser amount does not succeed in our proposed design.

### F. New set of Keys for every transaction

The keys are generated from random number *R* (generated by the MNO). The *R* acts as a seed for all keys. As *R* is fresh for every transaction, therefore the keys are also new in each transaction.

### G. Non-repudiation of transaction messages

The transaction result messages (message 23, message 25) are digitally signed. In case of any dispute about the payment, the MNO is to honour message 23 as it contains the MNO's digital signature. The shopping detail is also digitally signed by the shop POS terminal so the shop has to honour the prices mentioned in this message. Therefore the customer is completely secured about the transaction.

### H. Securing long term secret

$K_p$ is the long term secret between MNO and shop POS terminal. In our protocol, $K_p$ is used with the least exposure (only once). The security policy of the MNO can define the update of this key after a defined interval.

## VIII. CONCLUSION

In this paper, we have proposed a transaction protocol that provides a secure and trusted communication channel to the communication parties. The proposed protocol was based on NFC Cloud Wallet model for secure cloud-based NFC transactions. The operations performed by the vendor's reader, an NFC enabled phone and the cloud provider (in this paper MNO) are provided and such operations are possible by the current state of the technology as most of these measures are already implemented to support other mechanisms. We considered the detailed execution of the protocol and we showed our protocol performs reliably in cloud-based NFC transaction architecture. In addition, this paper provides other related issues that are required to be explored in order to find out different possible roles, accesses and ownerships of involved parties in NFC ecosystem. The main advantage of this paper is to demonstrate another way of payment for all those people who do not have bank accounts. This way of making payments eases the process of purchasing for ordinary people as they only have to top up with their MNO without having to follow all the banking procedures.

From the business and customers' point of view, this method of payment is a micropayment, involving a very small amount for transactions (typically less than £50.00); therefore, people can only use this payment method to pay for cheap products. This has also prevented the emergence of such system, as the cost for individual transactions must be kept low which is impractical.

*A. Future work*

As a part of future work, a proof of concept implementation can be carried out in order to determine the reliability of the proposed protocol in terms of number of factors such as timing issues. This implementation refers to the performance domain of the proposed protocol which can be taken into the account to consider the performance of the protocol rather than its security that is discussed in this paper. The idea of the proposed protocol can also be extended to a multi-party protocol. Furthermore, other possible architectures in this area should be explored and defined in order to finalize the most reliable architecture for cloud-based NFC payment applications.